\documentclass[preprint2]{aastex}
\usepackage{lscape}
\begin{document}

\title{Searching for Planets in the Hyades IV: Differential Abundance Analysis 
of Hyades Dwarfs\footnote{Some data presented herein were obtained at the
W.M. Keck Observatory, which is operated as a scientific partnership among
the California Institute of Technology, the University of California and
the National Aeronautics and Space Administration. The Observatory was made
possible by the generous financial support of the W.M. Keck Foundation.}}

\author{Diane B. Paulson, Christopher Sneden}
\affil{Department of Astronomy, University of Texas, Austin, TX 78712}
\email{apodis@astro.as.utexas.edu, chris@verdi.as.utexas.edu}

\author{William D. Cochran}
\affil{McDonald Observatory, University of Texas, Austin, TX 78712}
\email{wdc@astro.as.utexas.edu}
\slugcomment{To appear in the June 2003 issue of The Astronomical Journal}

\begin{abstract}
We present a differential abundance analysis of Hyades F-K dwarfs in search for 
evidence of stellar enrichment from accreted hydrogen deficient disk material.
Metallicites and relative abundance ratios of several species have been 
determined. We derive a cluster mean [Fe/H] $=$ 0.13$\pm$0.01. Two stars show 
abundances $\sim$0.2 dex larger than the cluster mean. Additionally, one star, which was 
added by a recent study as a cluster member, shows significantly lower
abundances than the cluster mean. These three stars have questionable membership 
characteristics.
The remaining stars in the survey have an rms of 0.04 dex in the differential 
[Fe/H] values. The Hyades cluster members have apparently not been significantly chemically
enriched. The abundance ratios of Si, Ti, Na, Mg, Ca and Zn with respect to Fe 
are in their solar proportions.

\end{abstract}


\section{Introduction}
The proposition that extrasolar planet host stars tend to be metal rich has
implications for the planet formation community. This fact, as shown by 
e.g. Gonzalez (1997, 1998), Laughlin \& Adams (1997) and Jeffery et al. (1997),
\nocite{Go97,Go98, LaAd97, JeBaCh97} although interesting, has not yet been given 
a single, universally accepted theoretical explanation. Two possibilities are reviewed well by
\citet{MuCh02} and \citet{SmCuLa01}$-$ either planets form preferentially around stars which 
are intrinsically richer in heavy elements, or the over-abundance of
metals is due to hydrogen deficient protoplanetary debris enriching the 
stellar photosphere. Only stars which 
have shallow convection regions will show metal enrichment due
to accretion; when a star has a large convective region, the additional
metals will be diluted beyond detection. For F stars and earlier, 
enrichment can occur during the early life as a main-sequence star, however, 
for G stars, enrichment is thought to occur only if the accretion has 
taken place after the first 10 Myr of pre-main sequence evolution. At this time the convective region has decreased in 
size \citep{MuCh02, PiDeCo01, LaAd97}. Later type stars should show  no 
detectable enhancement, even if large amounts of material are accreted. 

Current searches for planets include stars with vastly different 
chemical and formation histories. 
One good way to test the enrichment theory is by observing a star cluster, whose
members were presumably formed from homogeneous material. 
The key is to look
for star-to-star differences in heavy element content. Stars showing higher amounts of 
metals would have had to be enriched in some way, notably from H deficient 
material being accreted onto the stellar atmosphere. Similar programs have utilized this concept of eliminating initial composition variable by performing
a differential abundance analysis of binary stars \citep{LaGo01, GrBoCl01}.
And recently, a program similar to this one has been undertaken by 
\citet{Fu02}.

Abundances of Hyades stars  have been determined by several 
groups over the past few decades. 
These studies have provided increasingly more accurate abundances, as 
atomic data have improved and as stellar atmosphere models have become closer
approximations to physical reality. \citet{VaMo99} 
review the abundance studies of the Hyades from \citet{CoWaWi65} through 
their own work. It is evident that the measurement of heavy elements
have been studied in A-F stars, but
a detailed analysis in the lower mass dwarfs is lacking. In addition, many of 
these studies only include 1-2 dozen stars. Conti et al. studied 
various elements in 10 Hyades stars. They were also  
interested in looking for star-to-star differences to determine if the 
protocluster nebula was homogeneous. This survey provided the 
first evidence that 
Li in the Hyades is not uniform, while the abundances of several other elements
were. To within their stated error bars, Conti et al. determined 
that the abundance of Hyades members is constant for all elements but Li. Later, 
chemical composition studies (excluding studies of only Fe and/or Li, which are 
more numerous) were completed in A-F stars by \citet{BoHeCo77}, 
\citet{BuCo89}, \citet{FrBo90}, 
\citet{GaReHe93}, \citet{TaSa97}, 
\citet{HuAl98}, \citet{BuCo00}, and \citet{TaKaTa98}. For lower mass stars, heavy element abundance
determinations were only completed by the following (again excluding those 
studies only of Fe 
and/or Li): \citet{CoWaWi65}, \citet{KiHi96}
and \citet{BoHeCo77}. Papers also instrumental to the metallicity 
determination of the Hyades cluster are \citet{Bo89}, \citet{BrLaTo80},
\citet{BoBu88}, \citet{CaCSCa85}, and \citet{ChCaSt71}.

\citet{PiDeCo01} have estimated the amount of Fe that must be accreted onto the stellar
photosphere in order to enrich the measured [Fe/H] of a solar-type star
appreciably.  
They derive that a quantity of 10 M$_{\oplus}$ of Fe (a rough upper
limit to the Fe core of Jupiter) would increase the [Fe/H] of a solar-type
star by 0.09 dex. Within the errors of stellar abundance analysis and atomic
data, variations of this magnitude within the cluster are detectable through 
a differential abundance analysis.

In this paper, one in a series exploring planets and planet formation in 
the Hyades cluster (e.g. \nocite{CoHaPa02, PaSaCo02, PaSaCo03}Cochran et al. 2002, Paulson et al. 2002, 2003) we present
abundance determinations for Fe, Si, Ti, Ca, Na, Mg and Zn (as well as
differential measurements for each of these elements) for a large
sample of Hyades members over a wide effective temperature range in search for 
evidence of stellar enrichment.

\section{Observations}
\subsection{Sample}
The planet search program, undertaken with the Keck I HIRES \citep{VoAlBi94}, contains 98 F-M dwarfs 
\citep{CoHaPa02}.  The present chemical composition analysis, which makes use of
these spectra, unfortunately, does not include all 98 stars. The M dwarfs and 
a few late K dwarfs (B-V  cutoff of 1.0)
were not analyzed due to the crowded spectra, inability to accurately place
the continuum, and the poorer S/N achieved for these stars.
At the beginning of the survey, we 
selected only stars which were thought to have no stellar companions. In a 
few cases, we did select 
wide binaries which were sufficiently separated in the sky so that we would 
have no contamination in the spectra from the companion. However, since that time, 
four stars in the original sample are now known to have non-planetary-mass companions 
(vB~5, vB~52, vB~17$-$ \citet{PaGhRe98} and vB 88). The discovery of these 
stellar companions does not exclude us from detecting planets, nor should it affect the 
overall abundance determinations in this paper. Stars showing linear trends
in radial velocity 
(perhaps additional unknown binaries) have therefore been included. Also included in 
this sample is one star, HD~14127, which has been monitored in the planet search
program
but which, we are now confident, is not a member. A second, HIP~13600, may 
also not be a member. These are discussed further in \S 3.6.
The final sample size for this abundance study is 55 stars.

\subsection{Spectra}
A full description of the observations can be found in \citet{CoHaPa02}.
All spectra were obtained at the Keck I telescope from 1996-2002. We have 
used HIRES with resolving power (R = $\Delta\lambda / \lambda$) nominally at 60,000. 
The signal-to-noise ratio (S/N) of each spectrum is 
typically 100-200 per pixel. The wavelength range of (3805-6188\AA) was
chosen so that all I$_{2}$ absorption lines are included for radial velocity 
measurements and so that the Ca II H\&K lines could be monitored for  stellar
activity. Unfortunately, this wavelength range excludes  the spectral lines of 
many interesting 
elements (e.g. Li).  The spectra used for abundance analysis are those taken as 
``template'' spectra for use in the radial velocity program. Thus, these 
spectra are free of I$_{2}$ absorption. Observations of hot stars are 
unnecessary due to the extreme lack of moisture at the Mauna Kea site. Therefore, the
telluric absorption will be very weak. All Keck spectra were reduced 
using standard IRAF\footnote{IRAF is distributed by the National Optical Astronomy Observatories,
    which are operated by the Association of Universities for Research
    in Astronomy, Inc., under cooperative agreement with the National
    Science Foundation.} packages. 

\section{Abundance Analysis}
The abundance analysis makes use of the current version of the LTE line analysis 
code MOOG \citep{Sn73}.
The linelist is compiled from three sources$-$ Fe I and II $gf$-values were 
derived internally (as described in \S 3.2) and the remaining parameters 
were obtained from either R. E. Luck (2003, private communication) or \citet{KuBe95}. 
 The final linelist is 
given in Table 1, and the sources for excitation potential ($\chi$) and 
oscillator strength ($gf$-value) are listed as well. In deriving [Fe/H]\footnote{[A/B] = log (A/B)$_{\star}$ $-$ log (A/B)$_{\odot}$} and 
[X/Fe] values, solar values were derived from a solar spectrum (of Ceres) taken
through HIRES. Originally, we adopted the values from \citet{GrSa98},
but we found a difference of 0.14 dex between our derived 
log~$\epsilon$(Fe)$_{\odot}$\footnote{where log~$\epsilon$(Fe)$_{\odot}$~=~
n$_{\rm Fe}$/n$_{\rm H}$ + 12.0} and that found by Grevesse \& Sauval. The
difference is primarily due to instrumental effects. Thus, in order to eliminate
instrumental uncertainties, we used log~$\epsilon$(Fe)$_{\odot}$ as derived 
from our solar spectrum. The values of log~$\epsilon$(X)$_{\odot}$ were also
in disagreement with Grevesse \& Sauval, so again, we chose to use the values 
derived from our solar spectrum.

 
\subsection{Stellar parameters}
We determined stellar parameters using the template spectra obtained from 
Keck HIRES. We normalized the continuum using 
IRAF. All equivalent widths (EW's) were measured by fitting a 
Gaussian to the observed line profiles, also using IRAF. 

The stellar models used were interpolated\footnote{Interpolation software
was kindly supplied by McWilliam (1995, private communication) and updated by
Ivans (2002, private communication).} atmosphere models (with no convective
overshoot) based on the 1995 version of ATLAS9 code \citep{CaGrKu97}.
The relevant stellar parameters$-$ effective temperature ($T_{\rm eff}$), gravity 
(log~$g$), and microturbulence ($\xi$), were determined in the following manner.
$T_{\rm eff}$ values were obtained by requiring that Fe abundances of individual
lines be independent of excitation
potential ($\chi$). Microturbulence was determined by forcing Fe abundances 
from individual Fe I lines to be independent of line strength.
Surface gravity was derived by requiring ionization equilibrium$-$ the Fe abundance
derived from Fe I lines must match that derived from Fe II lines. 
All stellar parameters are determined simultaneously,
with the only requirement being that log~$g$ is confined to the range
4.2 to 4.7, a reasonable range given the known cluster distance. Derived 
stellar parameters are listed in Table 2. The method by which we determine
stellar parameters also gives us the log~$\epsilon$(Fe) for each
star. We thus list [Fe/H] for each star in Table 3.

In our abundance computations, we chose a van der
Waals line damping parameter option with the \citet{Un55} approximation.
We also experimented with other damping enhancements recommended by
\citet{BlLySm95} and \citet{Ho71} to determine if one is significantly better 
for this set of data
than the others. All three damping options were tested using
both the curve of growth analysis and by comparing the lineshapes in a 
spectral region using a synthesis approach. We determined that neither of these 
enhancements to the Uns\"old approximation yields a better fit to the lines
chosen. The effect of damping was more  
apparent in the abundance analysis, where it was clear that the damping
parameter affected the cooler stars more than the warmer stars (yielding 
slightly different abundances along the main-sequence). 
In an absolute 
abundance analysis, the choice of different damping parameters does not seem to 
affect the final results. In an absolute abundance analysis, the comparison is 
made to the overall solar abundance, which changes only slightly 
with the different damping options (0.02 dex).
The effect of damping on a differential analysis becomes
slightly more pronounced. We compared the results of differential abundances of
two stars and the sun with 
varying damping parameters. The greatest difference was in vB 143 (compared to
the standard vB 153) which showed changes of 0.06 dex between the Holweger and 
the Blackwell et al. suggested enhancements.
The solar spectrum (also compared to vB 153) showed a difference of 0.04 dex
and vB 15 (compared to vB 153) showed 0.03 dex. There is no significant trend in these results with 
stellar temperature, indicating that 
the choice of damping should not adversely affect
the differential analysis. We note that \citet{PrNaCa00} also see 
inconsistencies between these damping enhancements. 

The choice of models with no convective overshoot was made both by taking into 
consideration the recommendations of \citet{CaGrKu97} and by empirical testing. We 
first used models with convective overshoot, but we found a significant linear
trend of increasing [Fe/H] with increasing $T_{\rm eff}$. Initially
we thought we were seeing the effects of uniform enrichment up the main 
sequence. However the majority of our program stars are G and K dwarfs. Thus, if
enrichment were uniform, there ought to be a plateau in the K and late G 
dwarfs with slight increase in early G and late F dwarfs. And, this is not 
what we were seeing. The other concern was that the slope was large (roughly
0.15 dex from F to K). So, we experimented with models with no convective
overshoot. The 
trend of abundance with $T_{\rm eff}$ disappeared by using these models. Thus, 
we decided to use models with no overshoot for the entire analysis.

For each star, we measured  the EW's of 
12$-$20 (presumably) unblended Fe I lines and 5-9 unblended Fe II lines 
in the region 4490$-$6175\AA. We preferentially chose lines redward of 4500\AA\ due to the extremely 
crowded spectra blueward of this cutoff. Where possible we tried to maintain 
only lines which were significantly redward of this. Continuum placement 
becomes difficult in the blue end of the spectrum. The number of lines available
varied according to the temperature of 
the star. Because the emphasis of this work is differential, 
absolute accuracy of elemental $gf$-values are of less importance. 
However, in order to obtain correct stellar parameters, we wanted to use 
accurate Fe $gf$-values. In the same way all EW's were measured, we varied the 
$gf$-values until the line abundances matched the solar model and solar 
abundances. We used the Kurucz solar atlas \citep{KuFuBr84} for this analysis.  

In December of 2001, C. Allende Prieto obtained spectra of several 
super-solar mass Hyades stars with McDonald
Observatory's 2.7m Harlan Smith telescope using the coud$\acute{e}$ echelle spectrograph 
(R=60,000 with S/N of about 300-500 per pixel at 5800\AA).
Of these stars, 11 were also observed in our sample at Keck, providing a 
comparison and check of measurements of stellar parameters
between the data sets. The stars observed at McDonald are typically higher
mass, and thus higher $v$sin$i$ than the stars chosen for planet search at 
Keck. So,
several  of the stars observed by C. Allende Prieto are not easily analyzed 
with the equivalent width method for determining abundances
using the chosen linelist. We do not want to compromise the
internal consistency of this analysis by adopting a separate linelist
for this additional set of data. Therefore, in this paper, we will only
discuss the analysis of the 11 stars common to both data sets.

We compare the stellar parameters derived for these 11 stars 
in Table 4. EW's measured from the McDonald spectra tend to be, on average, a 
few m\AA\ larger than the Keck spectra. This yields slightly 
higher 
(roughly 0.05 dex) abundances for [Fe/H]. This enhancement is completely due 
to the use of different spectrographs. In general, the stellar parameters we derive 
are consistent between the 2 data sets.

Our determination of $T_{\rm eff}$ also agrees well with
the $T_{\rm eff}$ derived for stars in common with \citet{APLa99} as shown in 
Figure 3, noting that
Allende Prieto \& Lambert list log~$T_{\rm eff}$ rounded to 2 decimal places.
Thus, in Figure 3, the converted $T_{\rm eff}$ from Allende Prieto \&
Lambert values appear to be in discrete units. The mean
difference is $\langle T_{\rm eff, this study} - T_{\rm eff, Allende-Prieto}\rangle
$=34.6~K with a standard deviation $\sigma$=67.9~K.

\subsection{[Fe/H] and [X/Fe]}
Absolute [Fe/H] abundances are derived during the process of determining stellar parameters.
Using the derived stellar atmosphere models and measured EW's of 
various atomic lines listed in Table 1, we determined
elemental abundances ([X/Fe]). Derived [Fe/H] and [X/Fe] values are listed for 
each star in Table 3. 

Our derived mean Hyades metallicity is $\langle$[Fe/H]$\rangle$ = 0.13$\pm$0.01 with $\sigma$=0.05, not including HD~14127. If we also exclude vB~1 and 
vB~2, we obtain $\langle$[Fe/H]$\rangle$ = 0.13$\pm$0.01, $\sigma$=0.04. This is
in agreement with various surveys of Hyades metallicity, e.g. 
$\langle$[Fe/H]$\rangle$ = 0.16$\pm$0.04 \citep{BoBeKi02}, 0.20$\pm$0.10
\citep{BrLaTo80}, 0.130$\pm$0.026 \citep{Bo89}, and 0.12$\pm$0.03
\citep{CaCSCa85}. 

We now consider the stars with [Fe/H] not within 1$\sigma$ of the cluster mean 
(these outliers can be seen in the 
differential [Fe/H] plot of Figure 2, which is further discussed in \S 3.3). 
The two outliers with $\Delta$[Fe/H] (and [Fe/H]) lower than the cluster 
mean (dashed line), HD~14127 and 
HIP~13600, are disregarded due to questionable membership, as discussed in 
\S 3.6.

One might assume that for the outliers with high [Fe/H] (and $\Delta$[Fe/H]) 
(vB~19, vB~1 and vB~2, as seen 
in Figure 2) we have determined an incorrect
set of model parameters$-$ that, perhaps, our determined $T_{\rm eff}$ are too 
high. However, changing the model 
parameters by reasonable amounts cannot solve the entire problem
of their high abundances. For vB~19, the determination of log~$g$ may be 
questionable due to the fact that we were only able to measure 4 Fe II lines in
the spectrum. If the gravity is questionable, then the other parameters 
may be off slightly as well. We have derived an $T_{\rm eff}$
which is roughly 150 K higher than \citet{APLa99} (hereafter APL99) derived. 
 We note that APL99 interpolate theoretical isochrones with observed
data from HIPPARCOS to get stellar parameters, which is a different
approach than what we use. Thus, some amount of
disagreement is understandable between our study and theirs.
To test our abundance determination, though, we force vB~19
to have APL99's derived temperature and rederive the other
stellar parameters. 
We find  $\xi$=1.0 km~s$^{-1}$ and log~$g$=3.9. Together, 
these new
parameters give an [Fe/H] of 0.16. This is within 1$\sigma$ of
the cluster mean, though still on the high end. However, the new gravity derived
is in strong disagreement with 4.40 as derived by APL99.
So, assuming that 
the disagreement in log~$g$ can be explained by EW measurement error, the
high abundance in vB~19 may be reduced to within 1$\sigma$ of the cluster mean. 
Thus, we do not feel strongly 
that vB~19 is, indeed, enhanced relative to the cluster.

Doing the same test for
vB~1, forcing a $T_{\rm eff}$ of 6165 K gives log~$g$ of 4.1 (in disagreement
with APL99's 4.51), $\xi$ of 0.8 km~s$^{-1}$, and
[Fe/H] of 0.23, still significantly higher than the cluster mean. For the
final case of vB~2, forcing the $T_{\rm eff}$ to be 5888 K gives log~$g$ of 4.1
(APL99 derived 4.40), $\xi$
of 0.6 km~s$^{-1}$, and [Fe/H] of 0.24 (also significantly higher than the
cluster mean). So, the high abundances of vB~1 and vB~2 cannot simply be
explained by poor choice of models unless the true model parameters are 
drastically
inconsistent with what we have measured. Additionally, 
in his initial analysis, \citet{Fu02} also sees an enrichment in these 
two stars. So,
either vB~1 and vB~2 are not members or they have been enriched relative to
the cluster mean. The membership of these stars is further discussed in \S
3.6.

\subsection{Differential [Fe/H] and [X/Fe]}
By employing a differential abundance analysis (e.g. Gray 1992\nocite{Gr92}), one removes the uncertainty
in the oscillator strengths of the lines, which are often poorly known. 
Therefore, to answer the question of whether we see enrichment of metals
within the Hyades cluster, we use a self-consistent, differential abundance
analysis to look for any star-to-star metallicity variations. We do not employ 
the method of also deriving differential stellar parameters as \citet{LaGo01}
do, because our $T_{\rm eff}$ range is too large. In doing a line-by-line differential analysis, we will be able 
to place upper limits on the amount
of H deficient debris that could have been accreted onto a star's photosphere
relative to the cluster mean. 

In order to get differential abundances ($\Delta$[Fe/H] and $\Delta$[X/Fe]) we 
subtract log~$\epsilon$(X) of each line in each star with the same line in a 
comparison star (we chose the K dwarf vB~153). Thus, $\Delta$[Fe/H] and $\Delta$[X/Fe] are
the means of the 
differences for all lines in a given star. The scatter is significantly 
reduced. This gives a more accurate relative abundance than we can obtain by 
just taking the mean of all lines and subtracting that value from solar (the 
values [Fe/H] and [X/Fe]). Differential abundance values are listed in Table 
5 and are plotted in Figures 2 and 4 for each species, with typical error
bars shown in the bottom right hand corner of each panel.
For a given element, we 
determine the star-to-star variations with a standard deviation about the 
mean $\sigma\le$ 0.05 dex and the standard deviation of the mean significantly 
lower (less than 0.01 dex). $\Delta$[Si/Fe], $\Delta$[Ti/Fe], $\Delta$[Na/Fe], 
$\Delta$[Mg/Fe] and $\Delta$[Zn/Fe] are
all fairly consistent with $\Delta$[Fe/H]$-$ they are constant along the 
main-sequence with small scatter. The linear least squares fits to these
data reveal the following relationships:\\
\indent $\Delta$[Si/Fe] = $-$0.106 + 0.109(B-V), \\
\indent $\Delta$[Ti/Fe] = $-$0.014 $-$ 0.040(B-V), \\
\indent $\Delta$[Na/Fe] = \ 0.087 $-$ 0.129(B-V), \\
\indent $\Delta$[Mg/Fe] = $-$0.025 $-$ 0.004(B-V) and \\
\indent $\Delta$[Zn/Fe] = \ 0.001 $-$ 0.067(B-V). \\
$\Delta$[Ca/Fe] has 
a significant trend in abundance with color. The value of $\Delta$[Ca/Fe] is zero 
at the same B-V as the comparison star. Therefore, we believe this trend is
primarily due to the fact that these lines begin to move off the linear 
part of the curve of growth (start becoming saturated) in cooler stars. For
the other elements, when a line had this type of behavior, we removed it from 
the list (we preferred to maintain one linelist for the entire sample of stars).
However, there was not a reasonable number of lines available for us to remove 
all lines of Ca which behave like this. The least squares fit for Ca is \\ 
\indent $\Delta$[Ca/Fe]= \ 0.274 $-$ 0.372(B-V).

In Figure 4, there are 2 severe outliers. In the $\Delta$[Na/Fe] plot, vB~26 
and vB~105 have extremely high $\Delta$[Na/Fe] compared to the cluster. 
The cause of this enhancement is saved for future study.

Finally, Table 6 lists cluster mean abundances for both absolute and 
differential analyses.

\subsection{$v$sin$i$}
We also determine the rotational velocity ($v$sin$i$) for all stars. 
$v$sin$i$, the 
instrumental profile (IP), the macroturbulence ($\zeta$), and the limb darkening
are combined 
to form a ``smoothing'' parameter. This smoothing parameter is convolved with an intensity profile for the star. Because we derive an intensity profile from the 
stellar models, in order to determine $v$sin$i$, we only need to 
determine these other  broadening parameters. 
We synthesized
5 Fe I lines in this 6150$-$6180\AA\ region. For each line, we know the
abundance from determination of the stellar parameters. We then fit a ``smoothing'' 
parameter to each line including calculated, measured or estimated 
values for each of the other broadening parameters. 
The IP is measured by fitting a Gaussian to the lines of 
the Thorium-Argon (ThAr) calibration lamp.
The FWHM, as measured from the ThAr
calibration spectra, varies from 0.0918 to 0.0921\AA\ from the redmost to the
bluemost lines in the chosen region. The synthesis code is insensitive to this 
small a change. So, we used 0.09\AA\ as the IP broadening.
We estimated $\zeta$ according to the \citet{SaOs97} estimates for
active stars and using B-V from APL99.  
The limb darkening coefficient is estimated from \citet{Gr92}. 
Using the individual abundances and the above smoothing parameters,
the only unknown left is $v$sin$i$. We took a 
mean of the $v$sin$i$ derived for each of the 5 lines to determine the 
overall $v$sin$i$ of the star. In this manner, we are able to determine 
$v$sin$i$ to about 0.7~km~s$^{-1}$.

The upper panel of Figure 5 
shows $v$sin$i$ versus B-V for our target stars, and individual measurements 
of $v$sin$i$ are listed in Table 2. We see the expected decrease of 
$v$sin$i$ with decreasing mass and the expected spread due to 
the sin$i$ ambiguity. To estimate the actual rotational velocity of a star
in the cluster based solely on color (or mass or $T_{\rm eff}$) we only need
to fit a function to the upper envelope of the $v$sin$i$ data. In Figure 5 
we have done so for 3 different functions $for\ our\ data\ alone$. 
The best fit for our data was
with a 5th order polynomial (solid curve). The second panel in Figure 5 shows
our $v$sin$i$ measurements and fits to our data along with $v$sin$i$ from \citet{BoRoCa02},
\citet{BeMaMe84} (with B-V values from SIMBAD\footnote{This research has made 
use of the SIMBAD database, operated at CDS, Strasbourg, France.}), 
and selected dwarfs from \citet{Kr65}. The $v$sin$i$
values taken from literature were measured in different ways and typically
they do not  
remove the, albeit small, contribution of macroturbulence.
Our fits (solid, dashed and dotted lines) do not take into 
account the $v$sin$i$ values from literature. These fits are extended to 
show that, while consistent with our data, they do not correctly
quantify the true rotational velocity ``upper envelope'' for all Hyades stars.

\subsection{Errors}
There are several sources of error when measuring stellar abundances, especially
if one is interested in absolute abundances. External errors such as 
uncertainties in atomic 
parameters, choice of model atmospheres and solar abundances, can cause 
significant errors in determined absolute
abundances, while these are minimized in differential abundance analysis. 
Internal errors, such as measurement of stellar absorption lines, continuum placement, and choice of stellar model parameters can be minimized and quantified 
to some degree. 
Typically, we can repeatably measure EW's to $\lesssim$ 1 m\AA. On 
average, an overestimation of a single line's EW by 1 m\AA\ will give a higher 
line abundance by 0.02 dex. 
Individual cases obviously will depend on the S/N of the spectra (i.e. spectra with 
higher levels of
noise will have larger errors in EW determination). Lines having 
noise that caused problems in line fitting (i.e. where a noise spike was 
present in the line or where the feature was difficult to discern from the 
noise in the continuum) were not included 
in the analysis. Continuum placement is also a (somewhat) unquantifiable 
source of error. However, as an example, we changed the order of the 
cubic spline fit to the continuum for a given spectrum and measured the 
EW from the same set of lines. A change from a third order to a fifth order 
spline gave a decrease in the EW of, on average, 0.4 m\AA. 
Additionally, we can quantify some internal errors, even though most of the 
fitting for the parameters ($T_{\rm eff}$, $\xi$, and log~$g$) are done by eye.
Table 7 shows the abundance dependencies on model parameters. For Mg 
abundance determination, we were only able to measure 1 stellar line, and for 
Zn and Na, 2, and Ca, only 3. The absolute abundances quoted here for these 
elements are much less certain than, say, Fe, which has significantly many 
more lines. For absolute [Fe/H], [Ti/H] and [Si/H], the $total$ error on an
individual measurement is about 
0.1 dex, which includes both internal and external errors. 
Differentially, the uncertainties are much smaller, on the order of 
0.05 dex. This is smaller due to the removal of certain external errors like
the atomic $gf$ parameter. 

\subsection{Notes on cluster membership}
The following 9 stars have at least one anomalous characteristic compared 
to other cluster members. The stellar characteristics of interest here are
metallicity, chromospheric activity level, photometry, proper motion and 
parallax. 

HD~14127 was included as a member by \citep{PeBrLe98}
(hereafter Pe98)  based on HIPPARCOS
observations. D. Latham (private communication, 1999) concluded that this star is not a member because it
has too high a Hipparcos distance and the photometry is below the main sequence.
In \citet{PaSaCo02} (hereafter Pa02), we showed that that this star does
have activity levels consistent with the age of the Hyades. However, 
this star's metallicity is 0.25 dex below the cluster mean. Thus, 
it is severely inconsistent with the Hyades.
It is our belief that this star is not a member of the Hyades.

HIP~13600 has slightly low abundances in all elements but Mg. In the activity
analysis, it was also an outlier, showing much lower activity levels than
expected for a Hyades member (Pa02). Again, this star was included by Pe98 but Latham
concludes that the photometry is below the main sequence and the Hipparcos
distance is too high. \citet{HoAg99} reject this star as a cluster member.
Thus, we consider HIP~13600 is a probable non-member.

vB~118, HD~26257, HD~35768, and HD~19902 all show low activity levels (Pa02), but
in this analysis, they all have consistent metallicities with the Hyades
cluster mean. Of these, Latham concludes that HD~26257, HD~35768 and HD~19902
are not members based on the same criterion as above. 
He agrees that vB~118 is a member. Pe98 includes all of these 
stars as members. HD~26257 and HD~35768 were rejected by Hoogerwerf \& Aguilar 
and HD~35768 was also rejected by \citet{DeHoDe01}. 
At this point, we still consider vB~118 to be a member. The others are 
considered to be probable non-members.

vB~12, also showing consistent abundances and photometry, radial velocity and
Hipparcos distances, shows slightly high activity levels. We believe
that this star is most likely a cluster member despite its anomalous activity
level.

vB~1 and vB~2 are the two stars in the sample which have significantly higher
abundances than the cluster mean. They both have consistent photometry, radial 
velocity and Hipparcos distances for membership. vB~1 has a slightly low 
activity level as compared to the cluster mean. 
\citet{DeHoDe01} find that these stars are non-members 
based on the proper motion and trigonometric parallax analyses of both
\citet{De99} and \citet{HoAg99}. The radial velocities of both of these stars 
is 31 km~s$^{-1}$, well within the range of the cluster (28~$-$~42~km~s$^{-1}$).
They have similar proper motions and are near one another in the cluster 
(the difference in right ascension is 6.39$^{s}$ and only 2$^{m}$3.6$^{s}$ in 
declination).   
Our differential radial velocity curves of these two stars (which, admittedly, 
only spans $\sim$5 years) do not reveal any linear trends suggesting 
a relationship between them; although, the possibility
still remains that they are or once were a wide binary pair.
Additionally, it is well known that nearby solar-type stars are generally 
of solar metallicity or lower, for example see the recent survey by
\citet{GaGo02}. So, it is unlikely that these stars happen to have similar 
super-solar metallicities, are quite close in proximity, and are just random field stars.  
We consider these stars to have questionable
membership, but at this time we are not able to make a more solid 
classification.

\section{Discussion}
It is apparent now that at the 1$\sigma$ (or 0.04 dex) level, most Hyades members are
uniform in composition. The abundances of several elements were measured to  
support this assertion. 
The implication of this result
is that because the Hyades members were formed from homogeneous material, 
if significant enrichment of photospheres occurred, 
we would be
able to see evidence of the enrichment. And, since we do not see variations
in measured abundances, significant enrichment has not occurred in these stars. 
Recalling the calculations of \citet{PiDeCo01}, enrichment of 10 M$_{\oplus}$ of Fe will increase the stellar [Fe/H] by 0.09 dex. We are able to rule out 
enrichment of this magnitude in our higher mass stars. 
We are able to scale this relation and determine that we still do not see 
enrichment at even the 7 M$_{\oplus}$ level. 

vB~1 and vB~2 are interesting stars. They are significantly enriched relative 
to the cluster mean. Early surveys have always included them as members, but 
recently, \citet{DeHoDe01} assert that they are non-members. The question 
remains, then, if the stars are enriched members or whether they are
interlopers. These two stars certainly merit significant further study.

When the search for planets concludes, we will
be able to say more firmly whether planets exist in the Hyades and if they have
migrated, we will be able to place firm upper limits on the amount of debris 
that
could have been cast onto the star during this process. For now,
we are only able to place upper limits on the possible
enrichment due to possible disks. Moreover, we are confident that the material
that formed member stars was, in fact, homogeneous.

\acknowledgments
We would like to thank Carlos Allende Prieto and David Yong for use of data 
obtained at McDonald Observatory. Additionally, DBP would like to thank 
Jennifer Simmerer for many useful discussions about stellar abundance analysis. 
This material is based upon work supported by the National Aeronautics and 
Space Administration under grant NAG5-9227 issued through the Office
of Space Science. CS is supported by NSF grant AST-9987162.


\clearpage
\newpage
\begin{figure}
\plotone{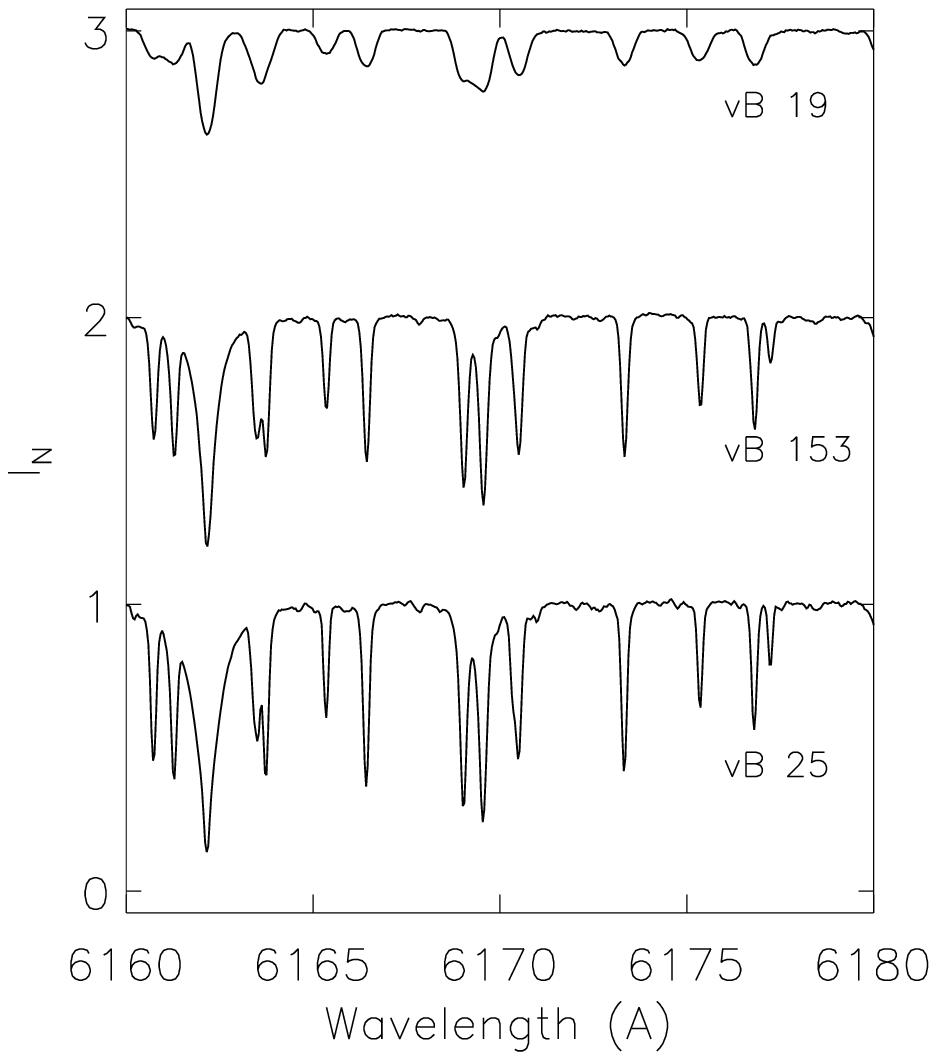}
\caption{Example spectra from our survey. We show here how some lines become
blended with increasing rotational velocity and increasing $T_{\rm eff}$. 
We have added 1 and 2 units, respectively, to the normalized intensity 
(I$_{N}$) of vB~153 and vB~19. vB~19 is our warmest star, vB~25 is our coolest
star and vB~153 is our comparison star.}
\end{figure}

\begin{figure}
\plotone{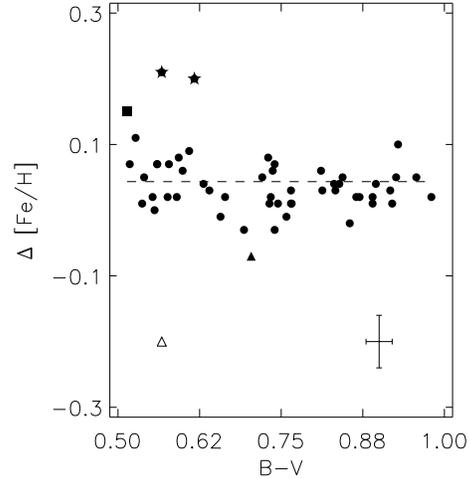}
\caption{Differential [Fe/H] of sample stars versus B-V. The differential 
comparison star is vB~153, as discussed in the text. Note that the 1 very low 
outlier is HD~14127 (open triangle) and 
HIP~13600 (filled triangle) is just slightly
lower in [Fe/H] than the other members.
vB~1, vB~2 (filled stars) and vB~19 (filled square) all are higher than other 
members. See \S3.6 for membership information.
The dashed line is the mean abundance level. A set of typical error bars for 
each measurement is shown in the bottom right hand corner of the plot.}
\end{figure}

\begin{figure}
\plotone{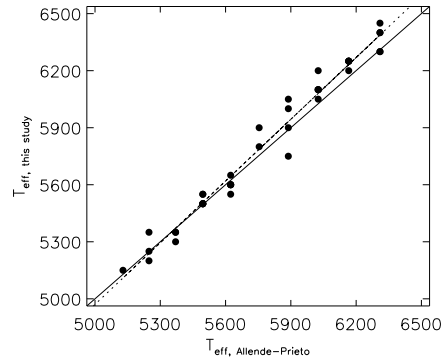}
\caption{A comparison of $T_{\rm eff}$ derived in this study versus those
derived by \citet{APLa99}. The solid line is 1:1, and the dashed line is
a least squares fit to the data.}
\end{figure}

\begin{figure}
\plotone{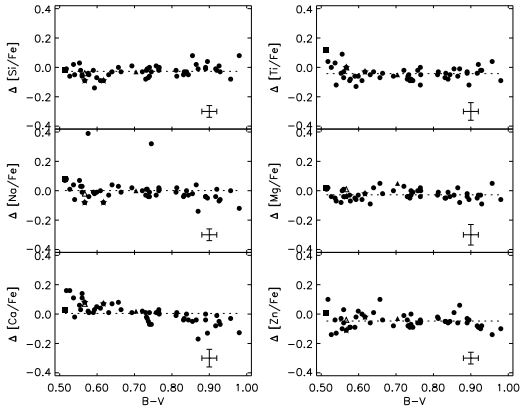}
\caption{Differential [X/Fe] of sample stars versus B-V.  The differential
comparison star is vB~153, as discussed in the text. The symbols are the same
as discussed in Figure 2.  Dashed lines are mean abundance levels. Typical error
bars are shown in each panel in the bottom right corner. }
\end{figure}

\begin{figure}
\plotone{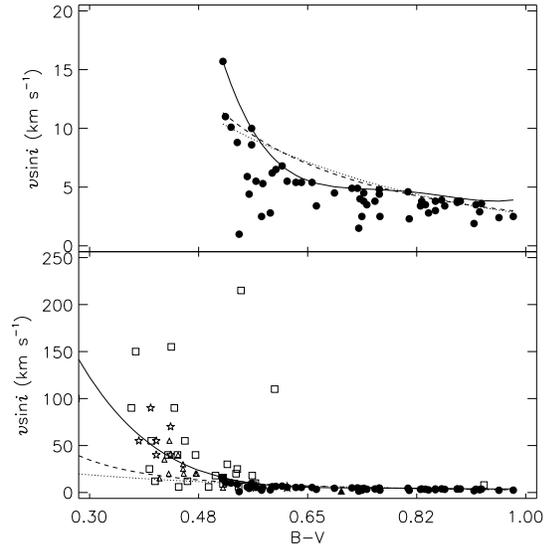}
\caption{$v$sin$i$ of observed Hyades stars (including the same possible 
non-members as discussed in Figure 2). The symbols are the same as in Figure 2.  The solid curve is a polynomial fit 
to the upper envelope of our data (filled circles) as discussed in the text, the dashed curve is a power law to our data only and the dotted curve is an exponential fit to 
our data, as well. The upper panel shows only our data whereas the bottom panel 
adds in $v$sin$i$ measurements from \citet{BoRoCa02} (open triangles), 
\citet{BeMaMe84} (open stars), and selected dwarfs from \citet{Kr65} (open
squares).}
\end{figure}

\clearpage
\newpage
\begin{deluxetable}{lcccc}
\tabletypesize{\scriptsize}
\tablecaption{Line List\label{tbl-1}}
\tablewidth{0pt}
\tablehead{
\colhead{Species} & \colhead{Wavelength} &
\colhead{$\chi$ (eV)} &
\colhead{log $gf$} &\colhead{Reference} }
\startdata
Na I & 6154.226   &  2.100  & -1.570& 1\\
Na I & 6160.747   &  2.100  & -1.270& 1\\
Mg I & 5711.088   &  4.346  & -1.833& 2\\
Si I & 5948.541   &  5.080  & -1.230& 1\\
Si I & 6125.021   &  5.610  & -1.513& 1\\
Si I & 6142.483   &  5.620  & -1.540& 1\\
Si I & 6145.016   &  5.610  & -1.479& 1\\
Ca I & 5260.387   &  2.521  & -1.719& 1\\
Ca I & 6166.439   &  2.520  & -1.142& 1\\
Ca I & 6169.042   &  2.520  & -0.797& 1\\
Ti I & 5922.110   &  1.046  & -1.410& 1\\
Ti I & 5937.811   &  1.066  & -1.834& 1\\
Ti I & 5941.752   &  1.053  & -1.454& 1\\
Ti I & 5953.162   &  1.887  & -0.273& 1\\
Ti I & 5965.828   &  1.879  & -0.353& 1\\
Ti I & 5978.543   &  1.873  & -0.440& 1\\
Ti I & 6064.629   &  1.046  & -1.888& 1\\
Ti I & 6126.217   &  1.066  & -1.369& 1 \\
Fe I & 5322.041   &  2.279  & -2.840& 2,3\\
Fe I & 5811.919   &  4.143  & -2.430& 2,3\\
Fe I & 5853.161   &  1.485  & -5.280& 2,3\\
Fe I & 5855.086   &  4.608  & -1.600& 2,3 \\
Fe I & 5856.096   &  4.295  & -1.640& 2,3 \\
Fe I & 5858.785   &  4.221  & -2.260& 2,3 \\
Fe I & 5927.797   &  4.652  & -1.090& 2,3 \\
Fe I & 5933.803   &  4.639  & -2.230& 2,3 \\
Fe I & 5940.997   &  4.178  & -2.150& 2,3 \\
Fe I & 5956.706   &  0.859  & -4.605& 2,3 \\
Fe I & 5969.578   &  4.283  & -2.730& 2,3 \\
Fe I & 6019.364   &  3.573  & -3.360& 2,3 \\
Fe I & 6027.051   &  4.076  & -1.090& 2,3 \\
Fe I & 6054.080   &  4.372  & -2.310& 2,3 \\
Fe I & 6105.130   &  4.549  & -2.050& 2,3 \\
Fe I & 6151.618   &  2.176  & -3.290& 2,3 \\
Fe I & 6157.728   &  4.076  & -1.110& 2,3 \\
Fe I & 6159.380   &  4.608  & -1.970& 2,3 \\
Fe I & 6165.360   &  4.142  & -1.470& 2,3 \\
Fe I & 6173.336   &  2.223  & -2.880& 2,3 \\
Fe II&  4491.407  &   2.855 &  -2.490&2,3 \\
Fe II&  4508.290  &   2.856 &  -2.310&2,3 \\
Fe II& 4620.520   &  2.828  & -3.230 &2,3\\
Fe II&  5197.559  &   3.230 &  -2.250&2,3 \\
Fe II&  5264.810  &   3.231 &  -3.150&2,3 \\
Fe II&  5325.559  &   3.221 &  -3.170&2,3 \\
Fe II&  5414.046  &   3.221 &  -3.620&2,3 \\
Fe II&  5425.247  &   3.199 &  -3.210&2,3 \\
Fe II&  6149.246  &   3.889 &  -2.720&2,3\\
Zn I & 4722.153   &  4.030  & -0.338&2\\
Zn I & 4810.528   &  4.078  & -0.137&2\\
\enddata
\tablerefs{
(1) Provided by E. Luck (2003, private communication)- compilation from 
various sources.\\
(2) From Kurucz Atomic Line Database \citep{KuBe95}.\\
(3) log $gf$-values derived from our solar spectrum and $\chi$ from \citet{KuBe95}.}
\end{deluxetable}

\begin{deluxetable}{lccccccc}
\tabletypesize{\scriptsize}
\tablecaption{Stellar Parameters\label{tbl-2}
}
\tablewidth{0pt}
\tablehead{
\colhead{HD} & \colhead{Other Name} &
\colhead{B-V (1)} &
\colhead{T$_{eff}$ (K)} &\colhead{log $g$ (cm s$^{-2}$)}&
\colhead{$\xi$ (km s$^{-1}$)} & \colhead{$\zeta$ (km s$^{-1}$) (2)}&\colhead{$v$sin$i$ (km s$^{-1}$)}
}
\startdata
26784& vB 19   &0.51&6450&4.2&1.1&4.89&15.7\\
27808& vB 48   &0.52&6400&4.3&1.0&4.86&11.0\\
30809& vB 143   &0.53&6400&4.2&0.9&4.79&10.1\\
28205& vB 65   &0.54&6250&4.3&1.0&4.71&8.8\\
28635& vB 88   &0.54&6250&4.3&0.8&4.69&1.0\\
26257& HIP 19386 &0.55&6300&4.3&1.0&4.59&5.9\\
35768& HIP 25639 &0.56&6300&4.3&1.0&4.57&4.4\\
27406& vB 31   &0.56&6200&4.3&1.0&4.54&10.0\\
28237& vB 66   &0.56&6250&4.3&0.7&4.54&8.6\\
20430& vB 1   &0.57&6250&4.4&0.8&4.49& 5.5\\
29419& vB 105   &0.58&6100&4.4&0.8&4.42&2.5\\
30589& vB 118   &0.58&6100&4.4&0.8&4.40&5.3\\
27835& vB 49   &0.59&6050&4.4&0.8&4.31&2.8\\
25825& vB 10   &0.59&6100&4.5&0.7&4.29&6.2\\
27859& vB 52   &0.60&6050&4.4&0.5&4.24&6.5\\
28344& vB 73   &0.61&6000&4.4&0.6&4.17&6.8\\
20439& vB 2   &0.62&6050&4.4&0.6&4.11&5.5\\
28992& vB 97   &0.63&5900&4.4&0.8&4.00&5.4\\
26767& vB 18   &0.64&5900&4.4&0.8&3.94&5.4\\
26736& vB 15   &0.66&5750&4.4&0.7&3.80&5.4\\
28099& vB 64   &0.66&5800&4.4&0.7&3.75&3.4\\
26756& vB 17   &0.69&5650&4.5&0.8&3.53&4.5\\
 & HIP 13600 &0.70&5600&4.5&0.6&3.44&1.8\\
27282& vB 27   &0.72&5600&4.5&0.7&3.32&4.9\\
240648& HIP 23750 &0.73&5600&4.5&0.7&3.25&4.9\\
19902& HIP 14976 &0.73&5600&4.5&0.8&3.23&1.5\\
28593& vB 87   &0.73&5550&4.5&0.8&3.22&4.0\\
31609& vB 127   &0.74&5550&4.5&0.6&3.19&2.5\\
26015B  & vB 12   &0.74&5250&4.5&0.8&3.17&4.5\\
28805& vB 92   &0.74&5500&4.5&0.7&3.17&3.8\\
27250& vB 26   &0.75&5550&4.5&0.8&3.13&3.5\\
27732& vB 42   &0.76&5500&4.5&0.8&3.03&3.8\\
32347& vB 187   &0.77&5500&4.5&0.8&2.98&4.4\\
242780& HIP 24923 &0.77&5500&4.5&0.7&2.98&4.8\\
283704& vB 76   &0.77&5500&4.5&0.8&2.97&2.5\\
284574& vB 109   &0.81&5350&4.5&0.8&2.63&4.6\\
284253& vB 21   &0.81&5350&4.5&0.5&2.62&2.3\\
285773& vB 79   &0.83&5300&4.5&0.5&2.48&3.4\\
30505& vB 116   &0.83&5300&4.6&0.8&2.46&3.8\\
28258& vB 178   &0.84&5350&4.6&0.7&2.42&3.5\\
  & vB 4   &0.84&5250&4.6&0.6&2.38&2.8\\
  & vB 153   &0.86&5200&4.6&0.7&2.30&3.8\\
27771& vB 46   &0.86&5200&4.6&0.8&2.30&3.0\\
28462& vB 180   &0.87&5250&4.6&1.0&2.22&3.9\\
29159& vB 99   &0.87&5000&4.6&0.5&2.18&3.4\\
28878& vB 93   &0.89&5150&4.6&0.7&2.03&3.8\\
285367& HIP 19098 &0.89&5150&4.6&0.8&2.03&3.7\\
285252& vB 7   &0.90&5050&4.6&0.8&1.99&3.8\\
  & vB 5   &0.92&5050&4.6&0.7&1.83&1.9\\
28977& vB 183   &0.92&5150&4.6&0.9&1.80&3.5\\
18632& HIP 13976 &0.93&5000&4.6&0.7&1.76&2.9\\
285830& vB 179   &0.93&5050&4.6&0.6&1.73&3.6\\
  & HIP 23312 &0.96&5100&4.6&0.7&1.52&2.4\\
285690& vB 25   &0.98&4900&4.6&0.8&1.35&2.5\\
14127& HIP 10672 &0.57&6200&4.4&0.8&4.49&6.3\\
\enddata
\tablenotetext{1}{B-V values taken from \citet{APLa99}}
\tablenotetext{2}{Macroturbulence derived from \citet{SaOs97}}
\end{deluxetable}

\clearpage
\begin{deluxetable}{lcccccccc}
\tabletypesize{\scriptsize}
\tablecaption{Elemental Abundances\label{tbl-3}
}
\tablewidth{0pt}
\tablehead{
\colhead{HD} & \colhead{Other Name} &
\colhead{[Fe/H]} & \colhead{[Na/Fe]} &
\colhead{[Mg/Fe]} & \colhead{[Si/Fe]} &
\colhead{[Ca/Fe]} & \colhead{[Ti/Fe]} & \colhead{[Zn/Fe]} }
\startdata
26784&vB 19 &0.23&0.07&-0.08&-0.01&0.08&0.08&-0.19\\
27808&vB 48 &0.15&0.07&0.00&0.06&0.15&0.15&0.11\\
30809&vB 143 &0.19&0.10&-0.06&0.02&0.29&0.15&-0.14\\
28205&vB 65 &0.10&0.03&-0.08&0.08&0.17&0.10&-0.05\\
28635&vB 88 &0.09&0.05&-0.05&0.05&0.11&0.01&-0.09\\
26257&HIP 19386&0.11&0.00&-0.11&0.09&0.12&0.03&-0.03\\
35768&HIP 25639&0.08&0.08&-0.02&0.05&0.12&0.12&-0.10\\
27406&vB 31 &0.15&0.05&-0.06&0.01&0.19&0.07&-0.06\\
28237&vB 66 &0.14&0.04&-0.01&0.02&0.22&0.03&0.04\\
20430&vB  1 &0.30&0.02&-0.05&0.00&0.16&0.07&-0.11\\
29419&vB 105 &0.13&-0.06&-0.05&0.01&0.11&-0.04&-0.07\\
30589&vB 118 &0.15&0.00&-0.08&0.02&0.09&0.00&-0.12\\
27835&vB 49 &0.09&0.00&-0.06&0.05&0.10&0.03&-0.08\\
25825&vB 10 &0.15&0.02&-0.07&-0.04&0.12&-0.03&-0.01\\
27859&vB 52 &0.14&0.01&-0.05&0.00&0.13&0.03&0.02\\
28344&vB 73 &0.18&0.01&-0.09&-0.03&0.11&0.01&0.00\\
20439&vB  2 &0.30&0.01&-0.06&-0.02&0.13&0.04&-0.03\\
28992&vB 97 &0.12&-0.01&-0.11&0.02&0.09&0.03&-0.05\\
26767&vB 18 &0.12&-0.01&-0.01&0.04&0.13&0.05&0.00\\
26736&vB 15 &0.09&0.05&0.01&0.06&0.14&0.00&0.08\\
28099&vB 64 &0.10&0.02&-0.04&0.05&0.11&0.04&-0.04\\
26756&vB 17 &0.06&0.01&-0.03&0.08&0.08&0.02&-0.08\\
      &HIP 13600&0.02&0.03&0.02&0.04&0.09&0.03&-0.04\\
27282&vB 27 &0.15&0.01&-0.01&0.03&0.08&0.05&-0.03\\
240648&HIP 23750&0.15&-0.01&-0.05&0.02&0.09&0.03&-0.02\\
19902&HIP 14976&0.09&0.00&-0.06&0.07&0.05&0.03&-0.08\\
28593&vB 87 &0.11&0.02&-0.07&0.06&0.03&0.02&-0.08\\
31609&vB 127 &0.15&0.01&-0.08&0.00&0.02&-0.01&-0.10\\
26015B      &vB 12 &0.17&-0.04&-0.07&0.00&-0.01&-0.01&-0.09\\
28805&vB 92 &0.08&-0.04&-0.05&0.06&0.06&-0.04&-0.02\\
27250&vB 26 &0.09&-0.02&0.00&0.09&0.04&0.06&-0.09\\
27732&vB 42 &0.10&-0.05&-0.04&0.06&0.08&-0.04&-0.06\\
32347&vB 187 &0.11&-0.05&-0.02&0.06&0.09&0.03&-0.05\\
242780&HIP 24923&0.12&0.02&-0.05&0.07&0.09&0.07&-0.04\\
283704&vB 76 &0.09&0.01&-0.04&0.07&0.06&0.03&-0.07\\
284574&vB 109 &0.13&0.01&-0.04&0.08&0.08&0.05&0.00\\
284253&vB 21 &0.14&0.02&-0.06&0.00&0.06&0.02&-0.05\\
285773&vB 79 &0.14&0.01&-0.09&0.02&0.02&0.00&-0.06\\
30505&vB 116 &0.13&-0.05&-0.09&0.05&0.02&0.03&-0.07\\
28258&vB 178 &0.15&-0.03&-0.06&0.02&0.05&-0.02&-0.09\\
      &vB  4 &0.15&-0.02&-0.09&0.02&0.02&0.03&-0.07\\
      &vB 153 &0.10&-0.03&-0.04&0.07&0.06&0.07&-0.01\\
27771&vB 46 &0.08&0.00&-0.07&0.14&0.00&0.05&-0.01\\
28462&vB 180 &0.08&-0.02&-0.02&0.09&0.01&0.03&-0.08\\
29159&vB 99 &0.09&0.04&-0.13&0.17&-0.12&0.05&0.03\\
28878&vB 93 &0.13&-0.05&-0.09&0.06&0.01&0.01&-0.05\\
285367&HIP 19098&0.11&-0.05&-0.08&0.05&0.03&0.02&-0.10\\
285252&vB  7 &0.15&-0.03&-0.11&0.10&-0.08&-0.06&-0.07\\
      &vB  5 &0.16&-0.08&-0.12&0.04&-0.05&-0.01&-0.13\\
28977&vB 183  &0.13&-0.01&-0.09&0.03&0.03&0.07&-0.14\\
18632&HIP 13976&0.18&-0.10&-0.10&0.06&-0.02&0.03&-0.15\\
285830&vB 179 &0.22&-0.08&-0.15&0.03&-0.03&0.03&-0.12\\
      &HIP 23312&0.18&-0.04&-0.11&-0.04&0.00&0.09&-0.20\\
285690&vB 25 &0.08&-0.08&-0.06&0.20&-0.03&0.02&-0.09\\
14127&HIP 10672&-0.12&-0.09&-0.01&0.02&0.14&0.10&-0.05\\
\enddata
\end{deluxetable}

\clearpage
\begin{deluxetable}{lccccccc}
\tabletypesize{\scriptsize}
\tablecaption{Comparison of Derived Parameters for Stars Common to Both
Samples\label{tbl-4}
}
\tablewidth{0pt}
\tablehead{
\colhead{Star}& \colhead{$\Delta$[Fe/H]} & \colhead{$\Delta T_{\rm eff}$ (K)}&
\colhead{$\Delta$ log $g$} & \colhead{$\Delta \xi$} (km s$^{-1}$)}

\startdata
vB 19&0.06&0&0&0.1\\
vB 10&0.03&0&0&0\\
vB 73&0.04&0&0&0\\
vB 118&0.06&50&0&0\\
vB 105&0.08&0&-0.1&-0.1\\
vB 66&0.03&0&0&0\\
vB 88&0.07&50&0&-0.1\\
vB 65&0.09&0&0&-0.1\\
vB 143&0.06&0&-0.1&-0.1\\
vB 48&0.01&0&0&0\\
vB 31&0.07&0&0&-0.1\\
\enddata
\end{deluxetable}

\clearpage
\begin{deluxetable}{lccccccccc}
\tabletypesize{\scriptsize}
\tablecaption{Differential Abundances\label{tbl-5}
}
\tablewidth{0pt}
\tablehead{
\colhead{HD} & \colhead{Other Name} &
\colhead{$\Delta$[Fe/H]} &
\colhead{$\Delta$[Na/Fe]} & \colhead{$\Delta$[Mg/Fe]}&
\colhead{$\Delta$[Si/Fe]} & \colhead{$\Delta$[Ca/Fe]}&
\colhead{$\Delta$[Ti/Fe]} & \colhead{$\Delta$[Zn/Fe]}}
\startdata
26784& vB 19     & 0.15& 0.08& 0.02&-0.02& 0.03& 0.12& 0.01\\
27808& vB 48     & 0.07& 0.08& 0.02&-0.01& 0.16& 0.04& 0.10\\
30809& vB 143    & 0.11& 0.01&-0.04&-0.06& 0.16& 0.00&-0.14\\
28205& vB 65     & 0.01& 0.04&-0.05& 0.02& 0.11& 0.03&-0.04\\
28635& vB 88     & 0.05&-0.06&-0.07&-0.05&-0.02&-0.12&-0.13\\
26257& HIP 19386 & 0.02& 0.07&-0.08& 0.03& 0.06&-0.04&-0.03\\
35768& HIP 25639 & 0.00& 0.03& 0.00&-0.02& 0.03& 0.09&-0.10\\
27406& vB 31     & 0.07& 0.03&-0.04&-0.05& 0.11&-0.03&-0.06\\
28237& vB 66     & 0.07&-0.01& 0.01&-0.06& 0.14&-0.07& 0.03\\
20430& vB 1      & 0.21&-0.08&-0.04&-0.09& 0.08& 0.00&-0.11\\
29419& vB 105    & 0.04& 0.39&-0.03&-0.04& 0.02&-0.09&-0.09\\
30589& vB 118    & 0.07&-0.03&-0.07&-0.05& 0.01&-0.08& 0.09\\
27835& vB 49     & 0.02&-0.01&-0.05&-0.02& 0.01&-0.06&-0.09\\
25825& vB 10     & 0.08&-0.02&-0.06&-0.14& 0.03&-0.13&-0.02\\
27859& vB 52     & 0.06&-0.01&-0.03&-0.07& 0.05&-0.06& 0.02\\
28344& vB 73     & 0.09& 0.00&-0.06&-0.09& 0.04&-0.09& 0.00\\
20439& vB 2      & 0.20&-0.08&-0.02&-0.09& 0.07&-0.03&-0.02\\
28992& vB 97     & 0.04&-0.03&-0.09&-0.05& 0.01&-0.05&-0.06\\
26767& vB 18     & 0.03& 0.04& 0.02&-0.03& 0.07&-0.03& 0.01\\
26736& vB 15     &-0.01& 0.03& 0.05& 0.00& 0.08&-0.07& 0.00\\
28099& vB 64     & 0.02&-0.01&-0.02&-0.03& 0.03&-0.04& 0.10\\
26756& vB 17     &-0.03& 0.02& 0.00& 0.01& 0.01&-0.06&-0.08\\
&HIP 13600&-0.07& 0.00& 0.05&-0.03& 0.02&-0.05&-0.03\\
27282& vB 27     & 0.05& 0.00& 0.03&-0.03& 0.02&-0.02&-0.01\\
240648& HIP 23750& 0.08&-0.03&-0.04&-0.08& 0.01&-0.08&-0.04\\
19902& HIP 14976 & 0.01& 0.01&-0.04&-0.02&-0.02&-0.06&-0.08\\
28593& vB 87     & 0.02& 0.01&-0.04&-0.01&-0.03&-0.05&-0.08\\
31609& vB 127    & 0.06&-0.04&-0.05&-0.07&-0.05&-0.09&-0.09\\
26015B & vB 12  & 0.07&-0.04&-0.03&-0.06&-0.07&-0.08&-0.07\\
28805& vB 92     &-0.03&-0.01& 0.00& 0.00& 0.01&-0.09& 0.00\\
27250& vB 26     & 0.01& 0.32&-0.01&-0.03&-0.07&-0.09&-0.09\\
27732& vB 42     &-0.03&-0.01& 0.00& 0.01& 0.01&-0.02&-0.06\\
32347& vB 187    & 0.01& 0.02& 0.02&-0.02& 0.03&-0.12&-0.04\\
242780& HIP 24923& 0.03& 0.00&-0.04&-0.01& 0.01& 0.00&-0.06\\
283704& vB 76    &-0.01& 0.01& 0.00& 0.00& 0.00&-0.05&-0.05\\
284574& vB 109   & 0.06&-0.01&-0.03&-0.02&-0.01&-0.05&-0.02\\
284253& vB 21    & 0.03& 0.02&-0.01&-0.06& 0.01&-0.03&-0.03\\
285773& vB 79    & 0.04&-0.04&-0.05&-0.05&-0.04&-0.07&-0.05\\
30505& vB 116    & 0.03&-0.03&-0.05&-0.03&-0.03&-0.04&-0.05\\
28258& vB 178    & 0.04& 0.00&-0.01&-0.03& 0.00&-0.08&-0.06\\
 & vB 4   & 0.05&-0.03& 0.04&-0.05& 0.04&-0.05&-0.05\\
 & vB 153 & \nodata & \nodata & \nodata & \nodata & \nodata & \nodata & \nodata \\
27771& vB 46     &-0.02&-0.02&-0.03& 0.08&-0.05&-0.01& 0.01\\
28462& vB 180    &-0.02& 0.04& 0.02& 0.02&-0.04&-0.06&-0.06\\
29159& vB 99     &-0.02&-0.14&-0.08&-0.01&-0.17&-0.01& 0.06\\
28878& vB 93     & 0.02&-0.04&-0.04&-0.01&-0.04&-0.04&-0.03\\
285367& HIP 19098&-0.01&-0.04&-0.02& 0.00& 0.00&-0.03&-0.06\\
285252& vB 7     & 0.04&-0.05&-0.06& 0.04&-0.13&-0.12&-0.04\\
  & vB 5  & 0.03&-0.04&-0.05&-0.01&-0.08&-0.04&-0.08\\
28977& vB 183    & 0.01& 0.01&-0.03&-0.02&-0.01& 0.02&-0.09\\
18632& HIP 13976 & 0.05&-0.07&-0.03& 0.01&-0.05&-0.02&-0.10\\
285830& vB 179   & 0.10&-0.06&-0.09&-0.03&-0.07&-0.02&-0.08\\
&HIP 23312& 0.05& 0.00& 0.05&-0.08&-0.03& 0.04&-0.14\\
285690& vB 25    & 0.02&-0.12&-0.06& 0.08&-0.12&-0.09&-0.10\\
14127& HIP 10672 &-0.20&-0.02& 0.01&-0.04& 0.07& 0.00&-0.04\\
\enddata
\end{deluxetable}

\clearpage
\begin{deluxetable}{lccccccc}
\tabletypesize{\scriptsize}
\tablecaption{Mean Cluster Abundances\label{tbl-6}}
\tablewidth{0pt}
\tablehead{
\colhead{}& \colhead{$\langle$[Fe/H]$\rangle$} &
\colhead{$\langle$[Na/Fe]$\rangle$}&
\colhead{$\langle$[Mg/Fe]$\rangle$}&
\colhead{$\langle$[Si/Fe]$\rangle$}&
\colhead{$\langle$[Ca/Fe]$\rangle$}&
\colhead{$\langle$[Ti/Fe]$\rangle$}&
\colhead{$\langle$[Zn/Fe]$\rangle$}}
\startdata
Absolute& 0.13&0.01&-0.06&0.05&0.07&0.03&-0.06\\
$\sigma$&0.05&0.09&0.04&0.05&0.07&0.05&0.06\\ \hline
Differential&0.04&-0.01&-0.03&-0.03&0.01&-0.04&-0.05\\
$\sigma$&0.05&0.06&0.04&0.04&0.07&0.05&0.05\\
\enddata
\end{deluxetable}

\clearpage
\begin{deluxetable}{lccccccccc}
\tabletypesize{\scriptsize}
\tablecaption{Abundance Dependencies on Model Parameters\label{tbl-7}
}
\tablewidth{0pt}
\tablehead{
\colhead{Example Star}& \colhead{Model Parameter}& \colhead{$\delta$[Fe/H]} & 
\colhead{$\delta$[Na/Fe]} & \colhead{$\delta$[Mg/Fe]} & 
\colhead{$\delta$[Si/Fe]} & \colhead{$\delta$[Ca/Fe]} & 
\colhead{$\delta$[Ti/Fe]} & \colhead{$\delta$[Zn/Fe]} } 

\startdata

vB 65 &T$_{\rm eff}\pm$50 & $\pm$0.04 & $\mp$0.02 & $\pm$0.01 & $\mp$0.02 & $\mp$0.01 & $\pm$0.01 & $\mp$0.01\\
($T_{\rm eff}$ = 6250)&log $g$$\pm$0.20 & $\pm$0.04 & $\mp$0.01 & $\mp$0.02 & $\mp$0.01 & $\mp$0.02 &$\mp$0.01 &0.00\\
&$\xi\pm$0.2 & $\mp$0.03 & $\pm$0.02 & 0.00 & $\pm$0.01 & $\mp$0.01 & $\mp$0.02 &$\mp$0.06 \\ \hline
vB 7 &T$_{\rm eff}\pm$50 & $\pm$0.01 & $\pm$0.03 & $\pm$0.01 & $\mp$0.03 & $\pm$0.03 &$\pm$0.05 & $\mp$0.03 \\
($T_{\rm eff}$ = 5050)&log $g$$\pm$0.2& $\pm$0.01 & $\mp$0.06 & $\mp$0.07 & $\pm$0.01 & $\mp$0.09 & $\mp$0.02 & $\mp$0.01 \\
&$\xi\pm$0.2 &$\mp$0.04 & $\pm$0.02 & $\mp$0.01 & $\pm$0.02 &0.00& $\mp$0.02 & $\mp$0.01 \\
\enddata
\end{deluxetable}

\end{document}